\documentstyle[twoside,fleqn,espcrc2]{article}
\input epsfig.sty
\newcommand{\be}[3]{\begin{equation}  \label{#1#2#3}}     
\newcommand{\ee}{ \end{equation}}
\newcommand{\ba}{\begin{array}}
\newcommand{\ea}{\end{array}}

\newcommand{\NP}[3]{{\em Nucl. Phys.}{ \bf B#1#2#3}}
\newcommand{\PRD}[2]{{\em Phys. Rev.}{ \bf D#1#2}}
\newcommand{\PRL}[2]{{\em Phys. Rev. Lett.}{ \bf #1#2}}
\newcommand{\MPLA}[1]{{\em Mod. Phys. Lett.}{ \bf A#1}}

\newcommand{\PL}[3]{{\em Phys. Lett.}{ \bf B#1#2#3}}

\newcommand{\AmS}{{\protect\the\textfont2
  A\kern-.1667em\lower.5ex\hbox{M}\kern-.125emS}}

\title{
{\small
\rightline{HUB-EP-97/03}
\rightline{hep-th/9701053}
\rightline{January 1997}
}
\vspace{1truecm}
Classical and quantum aspects of 4 dimensional 
black holes\thanks{Contribution for the ``30th Ahrenshoop
International Symposium on the Theory of Elementary Particles'', Buckow,
Germany, 27-31 Aug 1996}}

\author{Klaus Behrndt\address{Institut f\"ur Physik, 
        Humboldt Universit\"at, \\ 
        Invalidenstra\ss{}e 110, 10115 Berlin, Germany}%
         }

\begin{document}

\thispagestyle{empty}
\textheight 205mm
\topmargin -7mm

\begin{abstract}
In this talk I address some aspects in the recent developments for $N=2$
black holes in 4 dimensions. I restrict myself on axion-free solutions
that can classically be related to intersections of isotropic $D$- or
$M$-branes.  After reviewing of some classical properties I include
corrections coming from a general cubic prepotential. On the heterotic
side these are quantum corrections for these black hole solutions.
Finally, I discuss a microscopic interpretation of the entropy for the
extremal as well as near-extremal black hole.
\end{abstract}

\maketitle

\section{Introduction}
This paper is based on talks that I have given at the Buckow meeting
as well as two further occasions. The development was so rapid
in recent times, that I could not resist to add some further points that
have not been discussed in Buckow. I hope the reader will get a more
complete picture of this interesting topic.

\smallskip

In the last year a complete new picture of 4-dimensional (4d) black
holes appeared. The main ingredient came from the $D$-branes
\cite{da/le} and the discovery that they are black $p$-brane solutions
of the RR sector of string theory \cite{po}. Soon, it has been shown
that the 4d black holes can be seen as intersections or bound states
of these $D$-branes. The interesting point is, that it provides a
microscopic picture of the black holes in term of open string
states. Thus, one can give the Bekenstein-Hawking entropy a
statistical interpretation. The Hawking radiation appears as a
recombination of two open strings to a departing closed string. In
this picture seems to be no room for an information paradox.

\smallskip

In the first part I will introduce the basic properties of
supersymmetric 4d black holes and relate them to intersections of
$D$-branes. In the second part I am going to discuss corrections
coming from a general cubic prepotential.  On the heterotic side these
are quantum corrections and on the type II side these corrections are
related to the Calabi-Yau (CY) compactification.  Finally, using the
microscopic picture I describe the excitation and radiation of the
black hole and argue that the Bekenstein-Hawking entropy coincides
with statistical entropy, which counts the degeneracy of states.


\section{4d susy black holes}
Static supersymmetric black holes in 4 dimensions can be classified in
4 groups. Keeping only one gauge field and one scalar fields
these black holes are solutions of the effective action
\be010
 S = \int d^4 x \sqrt{|g|} \{R - 2 (\partial \phi)^2  + 
 e^{-2 a \phi} F^2 \} 
\ee
and are classified by the scalar coupling $a$. Generally, one can
find solutions for all $a$ \cite{gi/ma} but supersymmetry allows
only the discrete values $a=0, 1/\sqrt{3}, 1, \sqrt3$.
The metric is given by
\be020
ds^2 = -H^{-{2 \over 1 + a^2}} dt^2 + H^{2 \over 1 + a^2} d\vec{x}^2 
\ee
where $H$ is a harmonic function. For $a=0$ we have the standard selfdual
Reissner-Nordstr{\o}m solution and for $a=1$ it is the dilaton black hole
hole. For $a=1/\sqrt{3}$ or $a=\sqrt{3}$ one can get this black hole solution
from the compactification of the 5d Reissner-Nordstr{\o}m resp.\ as
Kaluza-Klein black hole. Let us look in more detail on
the case $a=0$. This solution can be written as
\be030
ds^2 = - {1 \over (1 + {q \over r})^2} dt^2 + (1+ {q \over r})^2
 \left[ dr^2 + r^2 d\Omega \right]
\ee
and for $r \rightarrow 0$  and $r=q e^{\eta / q}$ this solution becomes
\be040
ds^2 = - e^{ 2\eta / q} dt^2 + d\eta^2 \ + \ q^2 d\Omega \ .
\ee
This is a Bertotti-Robertson space ($AdS_2 \times S_2$) which is
non-singular (see figure 1).
\begin{figure}[t]
\begin{center}
\mbox{\epsfig{file=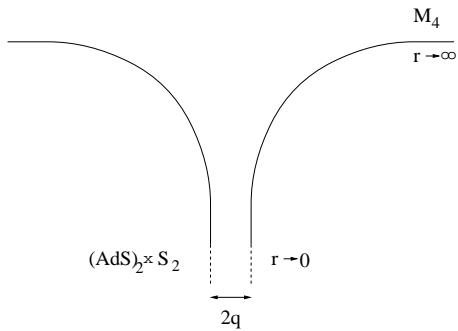, width=60mm}}
\end{center}
\vspace{-6mm}
\caption{The two asymptotic limits of the Reissner-Nordstr{\o}m
solution; flat space time at infinity and in the throat region
($r\rightarrow 0$) the Bertotti-Robertson space.}
\end{figure}

Generically, the Reissner-\-Nord\-str{\o}m solution breaks $1/2$ of
$N=2$ supersymmetries, but this asymptotic geometry restores all of
them.  Thus, this solution can be seen as a soliton that interpolates
between two maximal supersymmetric vacua, at infinity and on the
horizon. All other solutions with non-vanishing $a$ are singular at
$r=0$, but allow more unbroken supersymmetries. The $a=\sqrt{3}$
solution e.g.\ breaks only $1/2$ of $N=2$ supersymmetries in $D=10$.

In order to get the corresponding non-extreme solution we have to
replace
\cite{du/lu}
\be050
\ba{l}
dt^2 \rightarrow (1 - { \mu \over r }) dt^2 \  , \ 
dr^2 \rightarrow dr^2 \, / \, ( 1 - {\mu \over r}) \ ,\\
H = 1 + {q \over r} \rightarrow 1 + {\mu \sinh^2 \alpha \over r}
\ea
\ee
where $\mu$ is the non-extremality parameter and $\alpha$ is a boost
parameter. The extreme limit is: $\mu \rightarrow 0$, $\alpha
\rightarrow \infty$, but $\mu \sinh^2 \alpha = q$ (see also the
last paragraph). For all non-vanishing $a$ the horizon at $r=\mu$ touches
the singularity in the extreme case.  The situation becomes even worse 
if we would include angular momentum, where the singularity is
naked in the extreme limit. 


\section{Black holes as intersections of branes}
All these solutions have been generalised to the case of more gauge fields
and scalars. It is possible to see the $a=\sqrt{3}$ solution as a
fundamental state and all others as bound state of this fundamental one
\cite{ra}. In 10 dimensions this idea was confirmed by the interpretation of
these solutions as intersections of $D$-branes, which are $p$-brane
solutions of the RR sector of type II string theory. They are hypersurfaces
on which endpoints of open strings are attached \cite{po} and since the open
strings are fixed in the transverse directions (=Dirichlet boundary
conditions), these branes are called Dirichlet-branes (or $D$-branes).  This
was a major breakthrough in a new view, not only on these solutions but also
on strings or domain walls in 4 dimensions \cite{ber}.

Starting point was the reinterpretation of the G\"uven solutions
\cite{gue} as intersections of the 2-brane solution of the 11d
supergravity ($M$-2-brane) \cite{pa/to}. Soon after this paper
appeared all other solutions have been reinterpreted (see e.g.\
\cite{ts,be/be,ba/la}).  Let us start with some comments about these
brane solutions. First, they can have all dimensions, starting from
the (-1)-instanton until the complete 10d space, the 9-brane. The odd
branes are solutions of type IIB and the even branes of IIA
superstring theory. Furthermore, the (-1)-brane is defined in Euclidian
space time and the 7- \& 8-branes are not asymptotically flat (the
9-brane is the flat space time).

\begin{figure}[t]
\begin{center}
\mbox{\epsfig{file=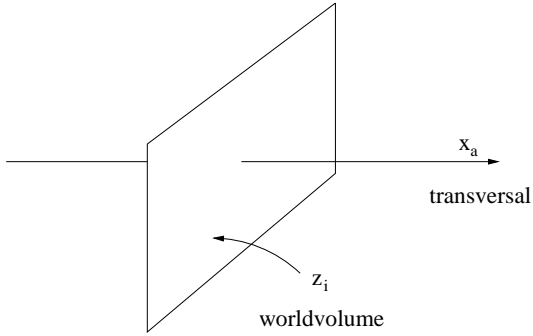, width=70mm}}
\end{center}
\vspace{-6mm}
\caption{A single isotropic branes as given \mbox{in (\ref{060}).}}
\end{figure}

Assuming that $z_i$ are the world-volume coordinates and
$x_m$ are the transversal directions these branes can be written as
(see figure 2)
\be060
\ba{l}
ds^2 = {1 \over \sqrt{H}}(-dt^2 + dz_i^2) + \sqrt{H} dx_m^2 \ ,\\
F = \left\{ \ba{l} dH \wedge dt \wedge .. \wedge dz_i \ , \ p \le 3 \\
   ^*dH \wedge dt \wedge .. \wedge dz_i \ , \ p \ge 3 \ea \right. \ , \\
e^{-2\phi} \sim H^{p-3 \over 2} \quad , \quad \partial^2 H(x) =0 \ .
\ea
\ee
These solutions are electric if $p<3$, magnetic if $p>3$ and for
$p=3$ we have the self-dual 3-brane. In addition
we have of course the brane solutions of the NS sector, the
fundamental string and the solitonic 5-brane. This completes the set of
charged brane solutions of the 10d effective string theory. As neutral
solutions we have the wave and the Taub-NUT soliton, which are pure 
gravitational solutions.

\smallskip

Since the type IIA theory follows by compactification of the 11d
supergravity, all even branes are compactified solutions too.  The 0-
\& 6-brane appear as KK solutions, whereas the 2- \& 4-brane are
compactified $M$-2- \& $M$-5-branes. Like the open string picture in 10d, we
have in 11d the possibility that an membrane can be open and end on a
5-brane \cite{to}. In difference, however, the open membrane has to end on a
5-brane whereas the open string can also be free. The $M$-brane solutions of
11d supergravity are 
\be070
\ba{l}
ds_2^2 = H^{- {2 \over 3}}(-dt^2 + dz_i^2) + H^{1 \over 3}dx_m^2 \ , \\
ds_5^2 = H^{- {1 \over 3}}(-dt^2 + dz_i^2) + H^{2 \over 3}dx_m^2  \ .
\ea
\ee
It is still an open question what is the 11d analogue of the $D$-8-branes
solution in 10d.

\begin{figure}[t]
\begin{center}
\mbox{\epsfig{file=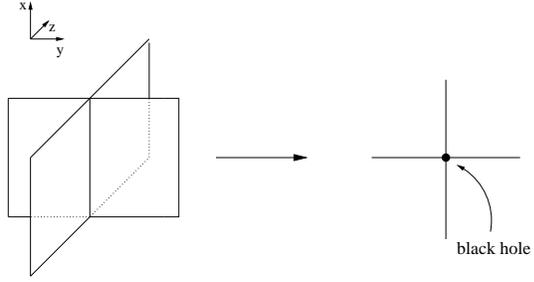, width=70mm}}
\end{center}
\vspace{-6mm}
\caption{Two 3-branes intersect over a string. Here $x$ is the common
worldvolume and $y$ and $z$ are relative transversal coordinates.
If we intersect two such objects we get a black hole at the intersection.}
\end{figure}

So far we have discussed single branes which yield after
compactification to 4d the $a=\sqrt{3}$ black holes. In order to get
also the other black hole solutions, we need to form bound states of
branes. These are states, where two or more
branes intersect each other perpendicularly (see figure 3). The
construction rules for these configurations can be found in
\cite{ts} or using $T$-duality in \cite{be/be}. 
E.g., two $D$-3-branes, which intersect over a string (=
common worldvolume) have the metric
\be080
\ba{l}
ds^2_{IIB} = {1 \over \sqrt{H_1 H_2}} du dv + \sqrt{H_1 H_2} dx_m^2 \\
\qquad + \sqrt{H_1 \over H_2 } (dx_6^2 + dx_7^2) + \sqrt{H_2 \over H_1 } 
(dx_8^2 + dx_9^2)
\ea 
\ee
where $u, v = z \pm t$ are the common world volume of the two 3-branes,
$x_m$ are 4 transversal coordinates and $x_6 .. x_9$ are the relative
transversal coordinates (they are worldvolume of one brane but
transversal for the other). Compactifying this configuration gives
however a singular solution in 4 dimensions, $G_{zz}$ and $G_{55}$ are
singular.  To keep these compactification radii finite we can make a
boost along $z$ and put a Taub-NUT soliton in the 4d transversal space,
i.e.
\be090
\ba{l}
du dv \rightarrow du dv + H_0 du^2 \\
dx_m dx_m \rightarrow {1 \over H_3} (dx_5 + \vec{V} d\vec{x})^2 + 
H_3 d\vec{x} d\vec{x} \ ,
\ea
\ee
where: $\epsilon_{ijk} \partial_j V_k = \partial_i H_3$ .
This is a non-singular type IIB solution. An analog solution on the
type IIA side can be uplifted to 11d. This solution is given
by three 5-branes which intersect over a string \cite{ts}
\be100
\ba{l}
 ds^2_{M} = (H_1 H_2 H_3)^{-{1 \over 3}} \left[ du dv + H_0 du^2 \right. \\
 \qquad + \left.  H_1 H_2 H_3 d\vec{x}^2 + H_A \omega^A \right]
\ea
\ee
where $\omega^A$ are three 2d line elements. The configuration with
coinciding harmonic functions has been discussed before in
\cite{pa/to}. These intersections contain all 4 classes of
black hole upon compactification to 4d. The number of non-trivial
harmonic functions determines the solution: the $a=0, {1 \over
\sqrt{3}} , 1 $ or $\sqrt{3}$ solution is given by an intersection
with 4, 3, 2 or 1 non-trivial harmonic functions (or branes).

These examples show also an interesting point concerning the
singularity of the 4d black holes. The $D$-3-brane as well as the
$M$-branes are non-singular objects \cite{gi/ho}. The compactification
of (\ref{080}) to 6d gives the self-dual string. Also this object is
non-singular - even behind the horizon is no singularity hidden, like
for the Reissner-Nordstr{\o}m black hole. A further compactification
to 4d gives us the singular $a=1$ black hole. Thus, the singularity of
this black hole is a consequence of the compactification. The same is
true for the other black holes. The $a=\sqrt{3}$ black hole is a
compactified single brane. So, we can take either the $D$-3-branes,
$M$-2- or the $M$-5-brane (\ref{070}) and compactify them to 4d and get
the $a=\sqrt{3}$ black hole. Also for the $a=1/ \sqrt{3}$ black holes
we can find intersections of non-singular branes. The magnetic case is
given by the intersection (\ref{100}) if we set $H_0=0$, i.e.\ an
intersection of three $M$-5-branes.  The electric solution corresponds
to an intersection of three $M$-2-branes
\be101
\ba{l}
 ds^2_{M} = (H_1 H_2 H_3)^{-{2 \over 3}} \left[ -dt^2 + \right.  \\
 \hfill + \left.  H_1 H_2 H_3 dx_m + H_A \omega^A \right]
\ea
\ee

where: $m=1..4$. Finally, the $a=0$ black hole can be obtained from
the intersection $3 \times 3 \times 3 \times 3$ or $2 \times 2 \times
5 \times 5$. But also the 5d black hole or black string, appearing
after compactification over the $\omega^A$-part in (\ref{101}) or
(\ref{100}), are non-singular.  Thus, all black holes, that we had
described in section 2, can be seen as compactification of non-singular
solutions.  The singularities are a consequence of the
compactification \cite{be/be1}. Of course, the Taub-NUT soliton and
boost are non-singular solutions too. The compactification of the
Taub-NUT soliton yields the known KK-monopole in 4d, which 
has also only a compactification singularity.

\smallskip

Also on the heterotic side, we can find a solution that includes all
four black hole types. This is the fundamental
string lying inside a solitonic 5-brane, again with a boost along the
fundamental string and a Taub-NUT soliton in transversal space. 
The metric is given by \cite{cv/ts}
\be102
\ba{l}
 ds_{het}^2={1 \over H_1} \left( du dv + H_0 du^2 \right) + dz_i^2 + \\
 \qquad +  H_2 \left({1 \over H_3} (dx_4 + \vec{V} d\vec{x})^2 + H_3 d\vec{x} 
 \right)
\ea 
\end{equation}
($\epsilon_{ijk} \partial_j V_k = \partial_i H_3$). Strictly speaking it is
not a genuine heterotic solution. Instead it is a solution of the NS-sector of
all string theories. The reduction to 4d can be done in two steps, first
going to 6d (compactifying $z_i$) and then to 4d. The internal space in
the second step is determined by two scalar fields $T$ and $U$ and in
addition we have the 4d dilaton $S$. We write the internal metric as
\be104
 G_{rs} = \pmatrix{ {H_0 \over H_1} & 0 \cr 0 & {H_2 \over H_3}}
  = {H_2 \over H_3} \pmatrix{ (\Im \,U)^2 & 0 \cr 0 & 1}
\end{equation}
and define the scalar fields by
\be106
 \begin{array}{l}
 S = i e^{-2 \phi} = i e^{-2 \hat{\phi}} \sqrt{|G_{rs}|} = 
  i \sqrt{{H_0 H_1 \over H_2 H_3}} \\
 T = i \sqrt{|G_{rs}|} = i  \sqrt{{H_0 H_2 \over H_1 H_3}} \  , \  
 U = i \sqrt{{H_0 H_3 \over H_1 H_2}}  
 \end{array}
\end{equation}
($\hat{\phi}$ is the 6d dilaton).
Note, that we are dealing with axion-free solutions, i.e.\ the real part 
of these fields vanish (see \cite{du/li}). 

\medskip

What about supersymmetry? Generically every brane breaks 1/2 of
supersymmetry. But all supersymmetries are broken if the dimension of
the relative transversal space is not 4 or 8 \cite{po/ch,ba/la} and
they do not fulfil the equations of motion \cite{be/be}. In
combination with the requirement to have an asymptotically flat
solution this restrict the maximal number of intersecting branes to
four. If one relax the restriction of asymptotical flatness or if one
is interested in compactifications to 3d or 2d one can have solutions
with more than 4 branes \cite{be/ro}. We see, many branes break many
supersymmetries. As consequence the renormalization behavior gets
worse. For enough unbroken supersymmetries and small curvature the
non-renormalization theorems prevent the solutions to get quantum
corrections. E.g.~in 4d, solutions of $N=4,8$ supersymmetric theories
are protected from quantum corrections but for solutions with only
$N=2$ supersymmetry we have to expect quantum corrections on the
one-loop and non-perturbative level.  This is especially relevant for
solutions with 3 or 4 intersecting branes.  In the $\sigma$-model
language on the heterotic side this means, that these background
fields are zeros only of the zero-loop-$\bar{\beta}$-functions but not
if we include the first genus. Also for solutions, where the curvature
is not under control we have to expect $\alpha'$-corrections. Even for
an $N=4$ theory these terms are not protected by a non-renormalization
theorem.  In the next paragraph we will discuss the way of how one can
address this question.


\section{Quantum aspects of $N=2$ black holes}
We are mainly interested in the compactified solution in 4
dimensions. All this solutions have a certain amount of supersymmetry
and can be embedded into known supergravity theories, which give us a
natural framework for the discussion of corrections.  These are not
only quantum correction. We have also corrections that are related to
the topology of the internal space, e.g.\ in the case of CY
compactifications. We will restrict ourselves on the $N=2$ case in
which the symplectic geometry gives us a powerful tool.

\medskip

Before we start we will fix our notations (see \cite{be/ca} and
refs. therein).  The $N=2$ supergravity includes one gravitational,
$n_{v}$ vector and $n_{h}$ hyper multiplets. In what follows we will
neglect the hyper multiplets, assuming that these fields are
constant. The bosonic $N=2$ action is given by 
\be110
\ba{rl}
S&\sim \int d^{4}x \sqrt{G} \{ R - 2 g_{A\bar{B}} \partial z^A \, \partial
\bar{z}^B + \\
& + {1 \over 4 } ( \Im {\cal N}_{IJ} F^I \cdot F^J + \Re {\cal
N}_{IJ} F^{I} \cdot {^{\star}F^{J}}) \} 
\ea
\ee 
where the gauge field part $F^I \cdot F^J \equiv F^I_{\mu\nu} F^{J \,
\mu \nu}$ and $I,J = 0,1 .... n_v$.  The complex scalar fields of the
vector multiplets $z^{A}$ ($A=1..n_{v}$) parameterize a special
K\"ahler manifold with the metric $g_{A\bar{B}} = \partial_{A}
\partial_{\bar{B}} K(z,\bar{z})$, where $K(z,\bar{z})$ is the K\"ahler
potential.  Both, the gauge field coupling as well as the K\"ahler
potential are given by the holomorphic prepotential $F(X)$
\be120 \ba{l} 
e^{-K} = i (\bar{X}^I F_I - X^I
\bar{F}_{I}) \\ {\cal N}_{IJ} = \bar{F}_{IJ} + 2i {(\Im F_{IL}) (\Im
F_{MJ}) X^{L} X^{M} \over (\Im F_{MN}) X^{M} X^{N}} 
\ea \ee 
with $F_{I} = {\partial F(X) \over \partial X^{I}}$ and $F_{MN} =
{\partial^{2} F(X)\over \partial X^{M} \partial X^{N}}$ (these are not
gauge field components).  The scalar fields $z^{A}$ are defined by
\be130 
z^{A} = {X^{A} \over X^{0}} 
\ee 
and for the prepotential we take the general cubic form 
\be140
F(X) = {1 \over 6} {C_{ABC} X^{A}
X^{B} X^{C} \over X^{0}} 
\ee
with general constant coefficients $C_{ABC}$. In type II
compactification these are the classical intersection numbers of the
Calabi Yau three fold.  On the heterotic side these coefficients
parameterize quantum corrections. The complete prepotential contains still
further terms, e.g.~we have omitted logarithmic and non-perturbative
corrections.  As long as we make sure that the scalars are large
($|z^A| \gg 1$) this is a good approximation.

\medskip

To find solutions to the above Lagrangian is in general very difficult.
However, restricting on the Reissner-Nordstr{\o}m case, i.e.~with 4
non-trivial harmonic functions, there exists a special limit in which
a solution can easily obtained. This is the double-extreme limit
\cite{ka/sh} where the scalar fields $z^A$ becomes constant everywhere.
On the heterotic side these are the fields as defined in (\ref{106}) $z^1 =
S$, $z^2 = T$ and $z^3 = U$ (if $n_v=3$). The values can be determined on
the horizon where the central charge is extremal \cite{fe/ka} and the
symplectic coordinates have to fulfil the constraint (in a certain basis,
see \cite{be/ca})
\be150
X^I - \bar{X}^I = i p^I \ , \ \partial_I F(X) - 
 \partial_{\bar{I}} F(\bar{X}) = i q_I
\ee
with $p^I$ and $q_I$ as the physical charges obtained by
\be160
q_I=\int_{S_{\infty}}G_{I} \quad , \quad
p^I = \int_{S_{\infty}} F^{I}
\ee
from the gauge fields $F^I_{\mu\nu}$ and $G_{I\, \mu\nu}= \Re N_{IJ}
F^{J}_{\mu\nu} - \Im N_{IJ}^{\ \ *}F^{J}_{\mu\nu}$. For the general cubic
prepotential these constraints have been solved for the axion-free case in
\cite{be/ca}. The solution with axions is not explicitely known - it is
given in terms of an algebraic constraint \cite{sh}. A special solution with
non-constant scalar fields could be obtained by replacing the charges in
(\ref{150}) by harmonic functions \cite{be}. This axion-free solution is
\be170
 \ba{l}
 ds^2 = - e^{-2 U} dt^2 + e^{2 U} d\vec{x} d\vec{x} \ , \\
 e^{2U} = \sqrt{H_0 \, {1 \over 6}C_{ABC} H^A H^B H^C} \ ,  \\ 
 z^A = i H_0 H^A e^{-2U} \ , \\
 F^A_{mn} = \epsilon_{mnp}\partial_p H^A \ , \  
 F_{0\; 0m} = \partial_m (H_0)^{-1}  \ .
\ea
\ee
Note that $F_{I \, \mu\nu} = {\cal N}_{IJ} F^J_{\; \mu\nu}$ and
$\Im N_{IJ} = {\cal N}_{IJ}$, the real part vanishes.
To be specific we choose for the harmonic functions
\be180
H^{A} = \sqrt{2}( h^{A} + {  p^{A} \over r} ) \  , \  
H_{0} = \sqrt{2}( h_{0} + { q_{0} \over r } )
\ee
where $h^{A}$, $h_{0}$ are constants which fixes the scalar fields at
infinity. This solution looks like a ``rotated version'' of the
solution with only 4 harmonic functions. We know from general duality
invariant $N=4$ solutions that they have a similar structure (see
e.g.\ the pure magnetic solution in \cite{be/ka1}).  But already the
special example discussed below tells us, that we cannot write this
black hole solution in terms of only 4 harmonic functions with {\em real}
coefficients. 

The black hole carries one electric and $n_v$ magnetic charges related
to the gauge fields $F^0_{0m}$ resp.\ $F^{A}_{mn}$.  To get the mass
we have to look on the asymptotic geometry. First, in order to have an
asymptotically flat Minkowski space we have the constraint
\be190
4 h_{0} {1 \over 6} C_{ABC} h^{A} h^{B} h^{C} = 1 \ .
\ee
Then, we get asymptotically
\be200
 e^{-2U} = 1 - {2 M \over r} \pm ....
\ee
and the mass is
\be210
 M = {q_{0} \over 4 h_{0}}  + {1 \over 2} p^{A} h_{0} C_{ABC} h^{B} h^{C} \ .
\ee
As expected our solution saturates the BPS bound
\be220
M^2 = |Z|^2_{\infty} = (q_0 X^0 - p^A F_A)^2_{\infty}
\ee
where the r.h.s.\ has to be calculated at spatial infinity ($e^U_{\infty} =1$).

\medskip

On the other side if we approach the horizon all constants $h^{A}$ and
$h_{0}$ drop out. The area of the horizon depends only on the conserved
charges $q_0,p^A$. Furthermore, if $q_{0} C_{ABC} p^{A} p^{B} p^{C}>0$ the
solution behaves smooth on the horizon and we find for the area and entropy
(${\cal S}$)
\be230
 A = 4 \, {\cal S} = 4 \pi \sqrt{4 q_{0} \, {1 \over 6} C_{ABC} 
 p^{A} p^{B} p^{C}}  \ .
\ee
If the charges and $h's$ are positive the area of the horizon defines
a lower bound for the mass. Minimizing the mass 
gives us the area of the horizon \cite{fe/ka}
\be240
4 \pi M^{2}|_{min.} = A \ .
\ee
In this case all scalar fields are constant and
\be250
h_{0} = {q_{0} \over c}  \qquad , \qquad h^{A} = {p^{A} \over c}
\ee
where $c^{4} = 4\, q_{0}{1 \over 6} C_{ABC} p^{A} p^{B} p^{C}$.  In
order to verify this, one has first to replace $h_0$ in (\ref{210}) by
using the constraints (\ref{190}). Then, using (\ref{250}) as ansatz
for $h^A$ one finds that: $\partial / \partial h^A M =0$ for all $A$
and thus, these values extremize the mass. On the other side if some of
the $p^A$ are negative, we can have massless black holes
and obviously (\ref{230}) cannot be the minimum.

For these moduli all scalars are constant, i.e. coincide with their
value on the horizon ($z^{A} \equiv z^{A} |_{hor.}$). By this
procedure we get the double extreme black hole \cite{ka/sh} and
taking this limit, our solution (\ref{170}) coincides with the
solution given in \cite{be/ca}. There is yet another way to look on
this extremization. The moduli fields are dynamical fields in $N=2$
supersymmetric gauge theories. Especially, the values at infinity
($h^A , h_0$) are not protected by a gauge symmetry. For a given model
there is no way to fix these values. Instead one could argue that the
model chooses those values for which the energy or ADM mass is
minimal, i.e.\ the double extreme case. This is the notion of
dynamical relaxation that has been introduced in \cite{re}.

\medskip

Let us end this paragraph with a discussion of the quantum corrections on
the heterotic side. To be explicit we consider the 3-moduli model with the
prepotential
\be260
F(X) = {X^1 X^2 X^3 \over X^0} + {1 \over 3} {(X^3)^3 \over X^0}
\ee
where the first term corresponds to the classical $STU$ model as described 
in paragraph 3 and the second is a quantum correction \cite{ha/mo}. 
For this model the metric is ($p^A \rightarrow {p^A \over \sqrt{2}}$, 
$q_0 \rightarrow {q_0 \over \sqrt{2}}$) 
\be270
\ba{l}
ds^2 = - e^{-2U} dt^2 + e^{2U} d\vec{x}^2 \\
e^{4U} = H_0 {1 \over 6} C_{ABC} H^A H^B H^C =  (1 + {q_0 \over r}) \cdot  \\
\cdot (1 + {p^3 \over r}) \left( (1 + {q^1 \over r})  (1 + {p^2 \over r}) 
+ {1 \over 3} (1 + {p^3 \over r})^2 \right) .
\ea
\ee
We see that the quantum term acts as a regulator. E.g.~if $q^1, p^2 < 0$
the singularity at $r=- q^1,p^2$ is regularized, which is the massless
black hole singularity. The solution behaves smooth even for vanishing
$q^1$ and $p^2$, which is classically not the case. But these questions
deserves further investigations, since near these points the validity of
this simple solution is questionable. 

The model gives a realistic picture as long as $H_0 \gg H^1 > H^2 > H^3
\gg 1$. The first inequality ensures that $S$, $T$ and $U$ are large and
all further corrections in the prepotential are sufficiently suppressed.
The last inequality suppressed higher curvature corrections ($\alpha'$
corrections), since the radius of the throat in the string-frame is
proportional to the magnetic charges (see figure 1). Near the horizon,
the above inequalities can directly be translated into relations between
the charges. If we have however negative charges, the harmonic functions
have zeros and near these points our approximation breaks down.

For the mass and entropy we get also additive corrections
\be280
\ba{l}
M = M_{cl} + {1 \over 4} p^3 \ , \\
{\cal S} = 2 \pi \sqrt{q_0 q^1 p^2 p^3 + {1 \over 3} q_0 (p^3)^3} \ .
\ea
\ee


\section{Microscopic entropy picture}
From the thermodynamics we know that the entropy should count the degeneracy
of states that build a physical system. For long time it has been an open
question what is this meaning for the Bekenstein-Hawking entropy of black
holes. What states do we have to count and where are they located? In the
last year a complete new picture of black holes emerged, the $D$-brane
picture \cite{st/va}, \cite{ca/ma}. The main idea is that the black holes
are wrapped $D$-branes (or $M$-branes) with open string (or membrane) states
on their surface. If we are far away it looks like the standard black hole,
but if we approach it, we see that the surface is not smooth, but consists of
clouds of open strings attached to the horizon (let us assume that the
horizon is non-singular). At the same time we will see that the black hole
(\ref{170}) is not really a point, but it is a string like object. This is
at least the case if we assume that the electric charge is large compared to
the magnetic ones, which in turn is a consistency relation to keep further
corrections under control. If the black hole is extremal (BPS solution) the
open strings move in one direction only.

\begin{figure}[t]
\begin{center}
\mbox{\epsfig{file=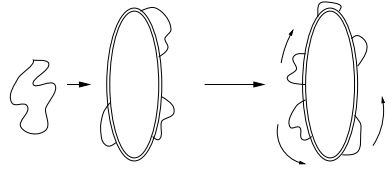, width=50mm}}
\end{center}
\vspace{-6mm}
\caption{All open strings are moving in
one direction along the closed string built out of
multiple layers of $D$-branes. If another closed
string falls in, it splits in two pieces one moving right
the other moving left.}
\end{figure}

\begin{figure}[t]
\begin{center}
\mbox{\epsfig{file=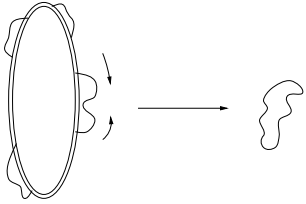, width=40mm}}
\end{center}
\vspace{-6mm}
\caption{Two open strings with opposite momenta can again
recombine and build a closed string. This is for $D$ branes
the qualitative picture of the black hole evaporation.}
\end{figure}

In this picture the excitation and the following evaporation of the
black hole has also a nice description (see figure 4 and 5). Taking a
closed string and throwing it into the black hole, it will break up
into two open strings, one will be right- and the other
left-moving. The black hole is excited or non-extremal. The part,
which moves in the opposite direction to all other open strings, will
eventually recombine with another open string and build a closed
string that can escape to infinity. In this picture seems to be no
place for an information loss, nothing is lost every state ``moves on
the horizon''.

A question remains however: why can the closed string not move behind the
horizon and is trapped there? In answering this question one could argue
that the discussion so far was a little bit naive. Let us repeat here
some arguments given in \cite{ho}. To talk about states makes sense only
in regions where the coupling to gravity is sufficiently small,
i.e.~where we can regard the (open string) states as small fluctuation
around flat space time. If the coupling increases a single state has no
sense. The back-reaction of gravity and all other fields will mix up all
states. If the coupling is large enough the complete system will collapse
into a black hole (or black string). A flat space description exists if
the gravitational field strength vanish for vanishing coupling. From
scaling arguments it follows that this is only the case for electric NS
solutions and the RR solutions, but not for magnetic NS solutions. Since
our black holes are dyonic only the RR charged solution has a flat space
description at weak coupling: in terms of $D$-branes and open strings.
Coming back to the question above, we can now argue that the described
$D$-brane picture just appears in the weak coupling regime. There, we do
not have a horizon, we have branes in flat space time and open strings
moving freely on these branes. Increasing now the gravitational coupling
the system collapse and our black hole (or black string) is created. In
the hope that the transition between both regions is smooth, we can
qualitatively describe the system in the $D$-brane picture. I.e.~we can
give an explanations of the Hawking radiation and also can count the
states relevant for the Bekenstein-Hawking entropy.

\smallskip

Let us now come to the question of state counting for our solution. This
has been done already for other solutions, which correspond to $K3$
compactification \cite{st/va} and the torus case \cite{ca/ma}. We will
concentrate here on the modifications that appear for the
CY-compactification.

Our solution can be obtained when we take 3 intersecting $M$-5-brane and
wrap them in the internal space. In the case that the internal space is a
torus this configuration is given in eq.~(\ref{100}). On the heterotic side
this corresponds to the classical case. For the state counting it is useful
first to compactify the 6d $\omega^A$ part, which gives us a string solution
in 5d. Wrapping this string around a further circle yields our black hole
in 4d. The states of the black hole can then be identified with the
states of the string in 5d. For the extremal case this state counting has
been done in \cite{be/mo} (see also \cite{ma} for the CY case of the type II
theory). Let us first repeat the arguments for the extremal case and later
discuss the non-extremal generalization.

\smallskip

The states that we have to count are open membranes attached to the
$M$-5-branes \cite{to}. After compactification, the open membranes become
open string states and since we want to consider first the extremal case
(BPS limit), they all travel in one direction. For the 5d string these
states are equivalent to momentum modes which yield the electric charge of
our solution (\ref{170}) upon compactification to 4 dimensions. Since the
$M$-5-brane can be wrapped many times around the CY-cycles, the 5d string
consist of many layers to which the states can be attached. This picture is
shown in figure 6.

\begin{figure}[t]
\begin{center}
\mbox{\epsfig{file=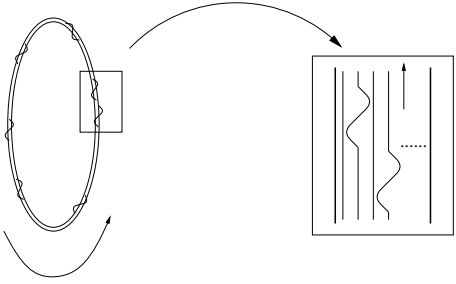,  width=60mm}}
\end{center}
\caption{This figure shows the 5-dimensional string with open string states
(=momentum modes) travelling in one direction (BPS limit).
In the simplest case the string consist of $p^1 p^2 p^3$ layers
}
\end{figure}

For the torus compactification a counting procedure has been proposed in
\cite{kl/ts} and we will adopt their arguments here. Before we start to
count let us describe the states shortly. The open membranes in 11
dimensions connect all three 5-branes. As consequence the open string states
in 5d consist of three open strings attached to the boundary at three
points.  They scale like $p^3$. What about other states? Membranes with less
than 3 boundaries are in general there, but they are subleading for large
charges (scale like $p^2$). On the other side membranes with more than 3
boundaries would destroy our picture, since the Bekenstein-Hawking entropy
scales like $p^3$. One can however argue that these membranes have to have
at least two boundaries on one 5-brane.  On the other side, every stretched
open membrane becomes massive and it tries to get the least mass, i.e.
either the two boundaries will recombine or they will move to a
self-intersection of the cycle and the state becomes massless there. For a
torus there are no self-intersections and thus only the open membranes with
3 boundaries are stable, branes with more boundaries will join all
boundaries at the same brane.  However, in the CY case we do have
self-intersections and we do have stable (massless) open membrane states
with more boundaries on the same 5-brane. But in every CY-3-fold at most
three 4-cycles (different of the same one) 
can intersect in one point and thus only open membranes with
3 boundaries can be stable and have the right scaling behavior ($\sim p^3$).
So, for our situation only the open membrane with 3 boundaries contribute to
the statistical entropy.

Next, how can we be sure that every state appears on the 5d string, that
nothing is hidden in the internal space? In wrapping the branes around
the internal cycles, we can shift the layers a little (see figure 7). Or
equivalently one could consider parallel (not coinciding) 5-branes. As
consequence, the open membranes are stretched at any point in the
internal space and they will move to the intersections, to become
massless there. The special case, that the 5-branes coincide in the
internal space does not change the number of states, only the location.

Now we can start to count these states and compare the Bekenstein-Hawking
entropy with the statistical entropy. For a large left-moving oscillator
number $N_L=q_0$ the statistical entropy of a string is given by 
\be290
S_{stat.} = \log d(N_L) = 2 \pi \sqrt{{1 \over 6} c_{eff} N_L} 
\ee
where $c_{eff}$ is the effective central charge which is given to leading
order by $c_{eff} = D_{eff}(1 + {1 \over 2})$ (bosonic plus fermionic
modes). Here, $D_{eff}$ is the effective number (= dimension) of oscillating
modes and it is given by the number of layers times the number of
oscillating modes for every layer. The number of layers is given by the
number of how many times we wrap a branes around the internal cycles. Lets
assume that we have three 4-cycles and
wrap the first brane $p^1$ times, the second brane $p^2$ and
the third brane $p^3$ times. Note, that we assume that the 5-branes have a
proper normalized charge in 10 dimensions. The charges in 4d comes from
multiple wrapping of the 5-brane. At the common intersection we have $p^1 p^2
p^3$ branes lying on top of each other (see figure 7).
\begin{figure}[t]
\begin{center}
\mbox{\epsfig{file=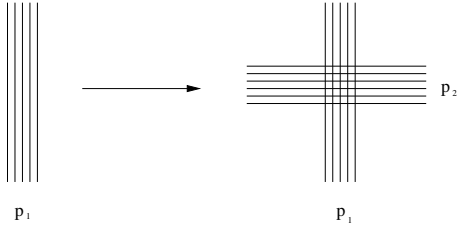, width=60mm}}
\end{center}
\vspace{-6mm}
\caption{Wrapping a brane $p^1$ times around a 4-cycle gives 
$p^1$ branes lying on top of each other. Intersecting two such 
4-cycles yield $p^1 p^2$ branes
that lying on the common intersection.
}
\end{figure}
Since each of these layers is part of one $M$-5-brane the oscillating
modes are oscillations of the 5-brane, which are the 4 transversal
oscillations.  As result we find that $D_{eff}= 4 \cdot p^1 p^2 p^3$.
This discussion can immediately generalized to the case that we have
$n_v$ 4-cycles around which we can wrap 5-branes. In this case we assume
that we have wrapped a 5-brane $p^A$ times around the $Ath$ 4-cycle ($A=1,
.., n_v$).  Assuming for the moment that we do not have
self-intersections, this can be written as $\frac{1}{6} C_{ABC} p^{A}
p^{B} p^{C}$, using that $C_{ABC}$ is symmetric and counts the number of
intersections of the 4-cycles. The resulting entropy is
\be300
S_{stat} = 
2 \pi \sqrt{q_{0}\, \frac{1}{6} C_{ABC} p^{A} p^{B} p^{C}}
\ee
and coincides with the Bekenstein-Hawking entropy (\ref{230})
and for the 3-moduli case with the entropy found in \cite{be/ka}.

What about self-intersections? At the beginning of our discussion we had
argued that the open membranes can sit with their boundaries on the same
5-brane, but only at self-intersections. In this case the endpoints are not
distinguishable and we have to divide out the permutation of the endpoints
sitting on the same brane, i.e.\ we have the combinatoric factor ${1 \over
6} = {1 \over 3!}$ if all boundaries are on the same brane. Hence, the
statistical entropy (\ref{300}) includes also the cases of
self-intersections.

\smallskip

Finally, let us come to the non-extremal case. In paragraph 2 we have seen
how we can make our solution non-extremal. For the entropy counting this
means that we also have to include right-moving modes. Let us explain
this. In order to make a brane intersection non-extremal we have to
introduce for every harmonic function a boost parameter (see second ref.\
of \cite{du/lu}). The appearance of these parameters can be understood as
follows. One can start with the Schwarzschild black hole and boost it
with finite velocity in one direction. Next, we $T$-dualize this
direction and get a non-extremal fundamental string. By $S$-duality we
convert it to a $D$-string and finally by further $T$-duality we can
create all other non-extreme $D$-branes. In order to get an intersection
we take, e.g., a non-extreme $NS$-5-brane. Again we boost this brane along
one of their world-volume coordinates and then by $S$- and $T$-duality we
can obtain all non-extreme intersections. This procedure can be
continued. So, for every brane we get a different boost parameter.

This procedure gives the right fields for toroidal compactifications
and the 4-d metric reads, see e.g.\ \cite{du/lu}
\be310
\ba{l}
 ds^2 = - e^{-2 U} f \, dt^2 + e^{2 U}({dr^2 \over 
  f } + r^2 d\Omega) , \\
 e^{2U} = \sqrt{\tilde{H}_0 \,\tilde{H}^1 \tilde{H}^2 
  \tilde{H}^3} \ ,  \\ 
\ea
\ee
where: $f= 1 - {\mu \over r}$, $\tilde{H}^A =\sqrt{2}( h^{A} +
{\tilde{p}^{A} \over r} )$ and $\tilde{p}^A = \mu \sinh^2
\alpha^A$. Now, the horizon is not $r=0$ but at $r=\mu$. We have to
keep in mind that this solution is not supersymmetric. This means that
it gets many more corrections.  We do not have anymore the tool of
special geometry, which we used in paragraph 4. But as long as we keep
$\mu$ very small compared to the charges and the boost parameter
large, we can see this solution as a small deviation from the extremal
solution. In the $D$-brane picture this means we have thrown only
very few states into a fat black hole.

The Hawking temperature can be obtained by going to the Euclidean
time. In order to avoid a conical singularity on the horizon, this
Euclidean time has to be periodic and as usual the periodicity
determines the temperature
\be320
0 \le t_e \le \beta = {1 \over T_H} \ .
\ee
Thus, we find for the Hawking temperature and the entropy,
which is again proportional to the area of the horizon
(setting $h^A = h_0 =1$)
\be330
\ba{l}
S =2 \pi \sqrt{ \hat{q}_0  \,  
 \hat{p}^1 \hat{p}^2 \hat{p}^3 } \\
T_H = {\mu \over 4 \pi \sqrt{\hat{q}_0  \,  
 \hat{p}^1 \hat{p}^2  \hat{p}^3 }} \ .
\ea
\ee
with: $\hat{p}^A = \tilde{p}^A + \mu = \mu \cosh^2 \alpha^A$.
The entropy has the same structure as before, only that it is 
expressed now in term $\hat{p}$. The extreme limit is given
by $\mu \rightarrow 0$ and $p = \tilde{p} = \hat{p}$. In this
case the Hawking temperature vanish, or equivalently the Euclidean
time becomes non-compact. 

Following the procedure of \cite{ho/st}, we can also give this entropy
a statistical interpretation. The main idea is, that the energy
of the in-following states is used to excite the wrapped branes,
brane - antibrane pairs are created. At the same time the in-falling
state (closed string) splits up into two pieces, one is left-moving
and the other right-moving (see figure 4). In order to motivate this idea
one looks on the physical charges, which are given by the non-extreme
field strength. In the non-extreme case they become
\be340
\ba{rcl}
q_0 & \rightarrow &  \mu \cosh \alpha_0 \, \sinh \alpha_0 = \\ & &
 = {\mu \over 4} ( e^{2 \alpha_0} - e^{-2 \alpha_0}) = q_0 - \bar{q}_0 \ , \\
p^A & \rightarrow &  \mu \cosh \alpha^A \, \sinh \alpha^A = \\ &&
= {\mu \over 4} ( e^{2 \alpha^A} - e^{-2 \alpha^A}) = p^A - \bar{p}^A \\
\ea
\ee
if: $q_0$ and $p^A$ are our extreme charges and $\bar{q}_0$ and
$\bar{p}^A$ are charges of anti-branes. Both numbers have to be integer
and counts the number branes and antibranes, or the left- and
right-moving momentum modes ($q_0 = N_L$ and $\bar{q}_0 = N_R$). Adopting
the notation of \cite{ho/st} we can replace in (\ref{330})
\be350
\ba{l}
\sqrt{\hat{q}} = \sqrt{N_L} + \sqrt{N_R} \ , \\
\sqrt{\hat{p}^A} = \sqrt{N_5^A} + \sqrt{N_{\bar{5}}^A} 
\ea
\ee
with: $N_5^A = p^A$ , $N_{\bar{5}}^A = \bar{p}^A$. So, in addition to momentum
modes travelling on the original branes we have now also to sum over
oscillations of the anti-brane part as well as right-moving modes.

Note, that the brane-antibrane picture is only very qualitative. It is
not a bound state, where we can separate only the anti-brane part. For a
vanishing brane part the antibrane part vanishes too. Thus, it is on
completely different footing as the non-threshold bound states of branes
discussed in recent literature (see e.g.\ \cite{ru/ts}).


\section{Conclusion}
We have reviewed some recent developments in 4d black hole solutions, that
can be embedded in $N=2$ supersymmetric theories. In comparison to $N=4$
solutions we had taken into account on the heterotic side quantum
correction and on the type II side correction, that come from the CY
compactification. Our model is described by a general cubic prepotential and
has one electric and $n_v$ magnetic charges. 

The 4d solution can be seen as a compactification of three $M$-5-branes.
If one first compactify over the CY-3-fold we get in 5d a string
solution, which is the common world-volume of all $M$-5-branes. The
number of wrapping one 5-brane around a 4-cycles gives us the magnetic
charge that corresponds to this 4-cycle. So the 5d string consist of many
branes lying on top of each other and along all of these layers we have
momentum modes travelling around the string. This gives us a microscopic
picture of the black hole and by counting of all states we were able to
determine the degeneracy to a given charge configuration, which coincides
with the statistical as well as Bekenstein-Hawking entropy. Making the
black hole non-extremal means, that the branes get excited, they split
into a brane and antibrane part and for the momentum modes we have left
as well as right moving modes. This microscopic picture has been
described in the last paragraph.

Still an open question is the inclusion of instanton corrections
on the type II side. These terms correspond to logarithmic corrections
in the prepotential. Especially in order to describe the massless
states these corrections are important. In our approach we could
suppress these terms by assuming that all charges are large, i.e.\ we
considered a large black hole.

\bigskip

\noindent 
{\bf Acknowledgements} \smallskip 

\noindent
I would like to thank Thomas Mohaupt for numerous discussions.
Work is supported by the DFG.

\end{document}